\documentclass[%
 aip,
cp,  % Conference Proceedings
 reprint,%
]{revtex4-2}

\usepackage{graphicx}% Include figure files
\usepackage{dcolumn}% Align table columns on decimal point
\usepackage{bm}% bold math
\usepackage[utf8]{inputenc}
\usepackage[T1]{fontenc}
\usepackage{mathptmx} 

\usepackage{graphicx}
\usepackage{amsmath}
\usepackage{amsfonts}
\usepackage{amssymb}
\usepackage{bm}
\usepackage{color}
\usepackage[x11names,dvipsnames]{xcolor}
\usepackage[colorlinks=true,allcolors=BlueViolet]{hyperref}

\newcommand\bsub{\begin{subequations}}
\newcommand\esub{\end{subequations}}
\newcommand{\diff}{\textrm{d}}
\DeclareMathOperator{\re}{Re}
\DeclareMathOperator{\im}{Im}

\newcommand\Sz{S^{(+)}_0}
\newcommand\Pz{P^{(+)}_0}
\newcommand\Dz{D^{(+)}_0}
\newcommand\Pm{P^{(+)}_{-1}}
\newcommand\Dm{D^{(+)}_{-1}}
\newcommand\Dmm{D^{(+)}_{-2}}
\newcommand\Pp{P^{(+)}_{1}}
\newcommand\Dp{D^{(+)}_{1}}
\newcommand\Dpp{D^{(+)}_{2}}

\newcommand\Pzs{P^{(+)*}_0}
\newcommand\Dzs{D^{(+)*}_0}
\newcommand\Pms{P^{(+)*}_{-1}}
\newcommand\Dms{D^{(+)*}_{-1}}
\newcommand\Dmms{D^{(+)*}_{-2}}
\newcommand\Pps{P^{(+)*}_{1}}
\newcommand\Dps{D^{(+)*}_{1}}
\newcommand\Dpps{D^{(+)*}_{2}}

\begin{document}

\title{Moments of Angular Distribution in Two Mesons Photoproduction}
\author{Vincent Mathieu}
\email{vmathieu@ucm.es}
\affiliation{Departamento de F\'isica Te\'orica, Universidad Complutense de Madrid, 28040 Madrid, Spain}

\begin{abstract}
The formalism devoted to the production of two pseudoscalar mesons with a linearly polarized photon beam has been detailed and illustrated in Phys. Rev. D100 (2019) 054017. This document reports the necessary formulas, without proof, to perform an analysis of the angular distribution of two mesons photoproduction. The relations to extract moments of the angular distribution are provided, as well as the relations between moments and partial waves for a system involving $S$, $P$ and $D$ waves. The expressions of the integrated beam asymmetry and the beam asymmetry along the $y$ axis in term of moments are mentioned. 
\end{abstract}

\maketitle

\section{Introduction}
With the 12 GeV upgrade, JLab enters a new era of meson spectroscopy. Several interesting problems in light meson spectroscopy involve resonances, such as scalar and hybrid mesons, that decay into two pseudoscalar particles. The identification of such mesons requires an analysis of the angular distribution of the decay products. This document aims at summarizing the key formulas to extract moments of the angular distribution of the two pseudoscalar mesons.  All formulas are extracted from Ref.~\cite{Mathieu:2019fts}. In this reference, the interested readers will find the derivations of the equations presented in this document, as well as an illustration of the moments and beam asymmetries. 

\begin{figure}[htb]
\begin{center}
\includegraphics[width=.45\linewidth]{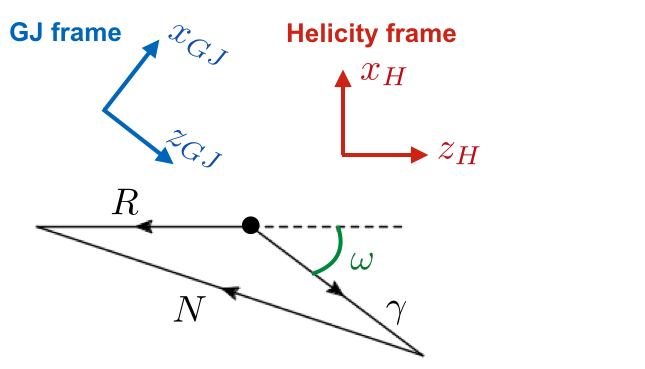}
\includegraphics[width=.54\linewidth]{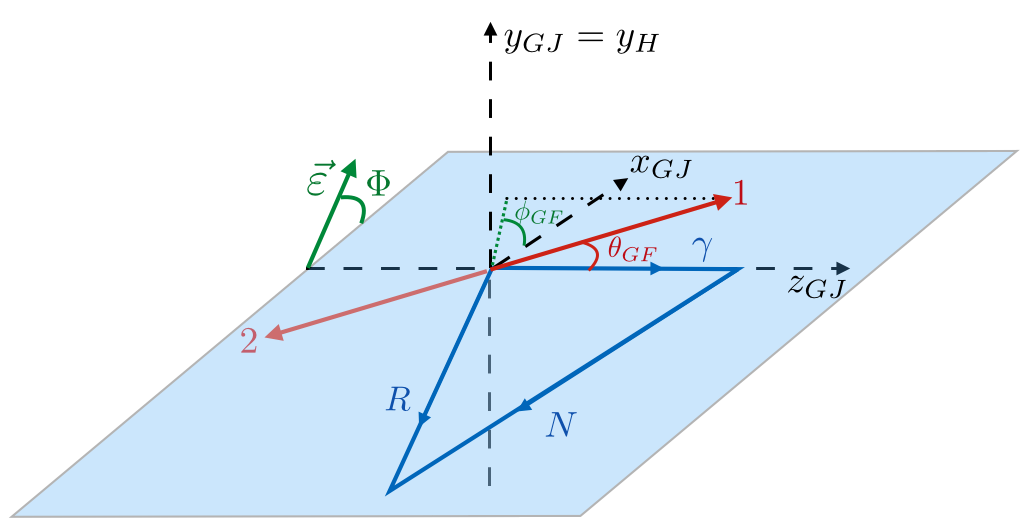}
\end{center}
\caption{\label{fig}{\it Left panel:} The reaction plane formed' by the beam $\gamma$, the nucleon target $N$ and the recoiling particle $R$. The $x$ and $z$ axes of the helicity and the GJ frames are indicated. The helicity frame is obtained from the GJ frame by a rotation of angle $\omega$ around the $y$ axis (perpendicular to the reaction plane). 
{\it Right panel,} The decay angles $\Omega_{GJ} = (\theta_{GF}, \phi_{GJ})$ of the meson $P_1$ in the GJ frame are indicated. The polarization vector $\varepsilon$ of the beam is also shown.}
\end{figure}

The photoproduction of two pseudoscalar mesons is decribed by the process
\begin{align} \label{eq:reaction}
\vec\gamma (\lambda, p_\gamma)\ p(\lambda_1, p_N) \to P_1 (p_1)\ P_2 (p_2)\ R (\lambda_2, p_R).
\end{align} 
In Eq.~\eqref{eq:reaction}, $P_{1,2} = \{\pi^\pm, K^\pm,\eta, \eta',  \ldots\}$ are the two pseudoscalar mesons and $R = \{p,n,\Delta^{++}, \ldots \}$ is the recoiling particle. The Mandelstam variables are the total energy squared $s = (p_\gamma + p_N)^2$, the momentum transferred between the target and the recoiling particle $t = (p_N-p_R)^2$, and the two mesons invariant mass squared $m^2 = (p_1+p_2)^2$. The dependence in the Mandelstam variables $s$, $t$ and $m$ will be implicit thorough this document as we are mainly focusing on the angular dependence. 
In the following equations, $m_N$ and $m_R$ are the masses of the nucleon target and recoiling particle respectively ; $m_{1,2}$ are the masses of the pseudoscalar mesons $P_{1,2}$.

The study of the angular distribution of the decay products is best perform in the center-of-mass of the two meson system. The beam, target and recoiling particle momenta form a plane, the reaction plane. The normal to this plane is conventionally the $y$ axis and several prescription exists for the $z$ axis. The two most commonly found choices are the helicity frame, in which the $z$ axis is the opposite of the recoiling particle's momentum; and the Gottfried-Jackson (GJ) frame, in which the $z$ axis is aligned with the beam momentum. The helicity and GJ frames are indicated on Fig.~\ref{fig}.
As it is well-known~\cite{Mathieu:2019fts, Mathieu:2018xyc} the the helicity (GJ) frame is equivalent to the $s$-channel ($t$-channel) frame when the target and the recoiling particles are not polarized. The $s$-channel being the center-of-mass of the direct reaction and the $t$-channel the center-of-mass of the crossed channel one.

The photon beam is linearly polarized with an angle $\Phi$ with respect to the reaction plane $xz$. The polarization angle $\Phi$ is extracted from the $y$ axis since the polarization vector is fixed in the lab frame and $\bm \epsilon(\Phi)\cdot\bm y = \sin\Phi$. The polarization angle $\Phi$ doesn't depend on the choice of the $z$ axis.

As illustrated on Fig.~\ref{fig}, the axes in these two frames are defined by (bold face indicates spacial three vectors)
\begin{align}
\bm z_{GJ} & = \phantom{-}  \frac{\bm p_\gamma}{|\bm p_\gamma|}, & 
\bm y_{GJ} & = \frac{\bm p_R \times \bm p_\gamma }{|\bm p_R \times \bm p_\gamma|}, &
\bm x_{GJ} & = \bm y_{GJ} \times \bm z_{GJ}, \\
\bm z_H & = - \frac{\bm p_R}{|\bm p_R|}, & 
\bm y_H & = \frac{\bm p_R \times \bm p_\gamma }{|\bm p_R \times \bm p_\gamma|}, &
\bm x_H & = \bm y_H \times \bm z_H.
\end{align} 
Given a choice of axes, $\Omega = (\theta,\phi)$ are the angles of one of the meson, say $P_1$. In the GJ frame, the angles are obtained from
\begin{align}
\cos \theta_{GJ} & = \frac{\bm p_1 \cdot \bm z_{GJ}}{ |\bm p_1|}, &
\cos \phi_{GJ} & = \frac{\bm y_{GJ} \cdot (\bm z_{GJ} \times \bm p_1)}{|\bm z_{GJ} \times \bm p_1|}, &
\sin \phi_{GJ} & = -\frac{\bm x_{GJ} \cdot (\bm z_{GJ} \times \bm p_1)}{|\bm z_{GJ} \times \bm p_1|}
\end{align}
The formula are identical in the helicity frame, with the axes replaced $\bm x_{GJ} \to \bm x_H$, etc.
These two frames are related by a rotation of $\omega$ around the $y$ axis as indicated in Fig.~\ref{fig}. The angle $\omega$ satisfies $\cos \omega = (\beta- \cos\theta_s)/(\beta  \cos\theta_s-1)$, with $\beta = \lambda^{1/2}(s,m_R^2, m^2) / (s-m_R^2 + m^2)$ and $\cos\theta_s$ is the cosine of the scattering angle between the nucleon target and the recoiling particle in the center-of-mass frame
\begin{align}
\cos\theta_s & = \frac{s^2 + s(2t - m_N^2-m_R^2-m^2) - m_N^2(m^2-m_R^2)}{\lambda^{1/2}(s,m_N^2,0) \lambda^{1/2}(s,m^2,m_R^2)}
\end{align}
The triangle function is $\lambda(a,b,c) = a^2+b^2+c^2- 2(a b + b c  + c a)$. 
At high energies, the crossing angle is simply $\cos \omega \simeq (m^2+t)/(m^2-t)$. 
The relation between the helicity and GJ angles are
\begin{align}
\cos \theta_{H} & = \cos \theta_{GJ} \cos\omega - \sin \theta_{GJ}  \cos \omega \sin \omega, &
\cot \phi_{H} & = \frac{1}{\sin \phi_{GJ}} \left[  \cos \phi_{GJ} \cos\omega + \cot \theta_{GJ} \sin \omega \right].
\end{align} 
The inverse relations are obtained by changing the sign of $\omega$.

%-------------------------------------------------------------------------------------------------
\section{Moments of Angular distribution}
The amplitudes for the reaction \eqref{eq:reaction} are denoted $A_{\lambda; \lambda_1\lambda_2} (\Omega)$. Since the formalism is equivalent in both the helicity and the GF frame, we will omit the frame index when there is no ambiguity. 
The $\Phi$-dependence of the intensity is encoded in the density matrix of the photon $\rho^\gamma$ and the differential cross section in photoproduction is
\begin{align}
I(\Omega,\Phi)&=  \frac{\diff\sigma}{\diff t  \diff m \diff\Omega\diff\Phi} 
 = \kappa
\sum_{ \substack{\lambda, \lambda' \\ \lambda_1,\lambda_2}} A_{\lambda; \lambda_1\lambda_2} (\Omega) \rho^\gamma_{\lambda\lambda'}(\Phi) A_{\lambda'; \lambda_1\lambda_2}^* (\Omega). 
\end{align}
We include all numerical factors in the phase space factor,%\footnote{The phase space factor is often absorbed in a redefinition of the amplitudes $\widehat T \equiv \sqrt{\kappa} T$ since it is numerically more stable to extract $\widehat T$ from data near the $\eta\pi^0$ threshold, where $\kappa \to 0$.}
\begin{align} \label{eq:pspace}
\kappa & = \frac{1}{(2\pi)^3} \frac{1}{4\pi} \frac{1}{2\pi} \frac{\lambda^{1/2}(m^2, m^2_1, m^2_2)}{16m(s-m^2_N)^2}\frac{1}{2}.
\end{align}

With a linearly polarized beam, the dependence on the polarization angle $\Phi$ can be made explicit. The photon density matrix doesn't depend on the choice of the $z$ axis and is thus equivalent in both helicity and GJ frames, $\rho^\gamma(\Phi)|_H = \rho^\gamma(\Phi)|_{GJ} = (1/2) \left[\sigma^0 -  \sigma^1 P_\gamma \cos 2 \Phi - \sigma^2 P_\gamma \sin 2 \Phi   \right]$, with $0 \leqslant P_\gamma \leqslant 1$, the degree of polarization. 
With the explicit dependence on the polarization angle, the intensity becomes:
\begin{align} \label{eq:int_pol}
I(\Omega,\Phi) & = I^0(\Omega) - P_\gamma I^1(\Omega) \cos 2 \Phi - P_\gamma I^2(\Omega) \sin 2 \Phi.
\end{align}

From the experimental intensity $I(\Omega,\Phi)$, one can extract the moments of the meson angular distribution. With a linearly polarized beam, the kinematical dependence in the polarized angle $\Phi$ in Eq.~\eqref{eq:int_pol} leads to naturally define unpolarized $H^0(LM)$ and polarized $H^{1,2}(LM)$ moments. They are unambiguously extracted from:
\bsub  \label{eq:int2mom2}
\begin{align} 
H^0(LM)  &= \frac{1}{2\pi}\, \int_0^{2\pi}\int_0^{2\pi} \int_{-1}^1 \  I(\Omega,\Phi) \, d^L_{M0}(\theta) \cos M\phi\ \diff(\cos \theta)  \diff \phi  \diff \Phi ,\\
H^1(LM)  &= \frac{1}{P_\gamma \pi} \int_0^{2\pi}\int_0^{2\pi} \int_{-1}^1 \   I(\Omega,\Phi) \, d^L_{M0}(\theta) \cos M\phi\, \cos 2 \Phi \ \diff(\cos \theta)  \diff \phi  \diff \Phi,\\
\im H^2(LM) & =  \frac{-1}{P_\gamma \pi} \int_0^{2\pi}\int_0^{2\pi} \int_{-1}^1 \  I(\Omega,\Phi)\, d^L_{M0}(\theta) \sin M\phi \, \sin 2 \Phi \ \diff(\cos \theta)  \diff \phi  \diff \Phi.
\end{align}\esub
Parity conservation restricts the independent moments to $0 \leqslant M \leqslant L$. The extra minus sign in the definition of $H^1(LM)$ ensures that $H^1(00)$ is positive for positive reflectivity waves. 

The three sets $\{ H^0(LM), H^1(LM), \im H^2(LM) \}$ are equivalent to the full intensity since the complete angular distribution can be recovered from Eq.~\eqref{eq:int_pol} and 
\begin{subequations}\label{eq:defHa}
\begin{align}
I^{0}(\Omega) & =\ \, \sum_{L,M\geqslant 0} \left(\frac{2L+1}{4\pi} \right) (2-\delta_{M,0}) H^{0}(LM) d^{L}_{M0} (\theta) \cos M\phi,  \\ 
I^{1} (\Omega)& = - \!\!\!\! \sum_{L,M\geqslant 0} \left(\frac{2L+1}{4\pi} \right) (2-\delta_{M,0}) H^{1}(LM) d^{L}_{M0} (\theta) \cos M\phi, 
\\ 
I^2 (\Omega)& = 2 \sum_{L,M>0} \left(\frac{2L+1}{4\pi} \right)  \im H^{2}(LM) d^{L}_{M0} (\theta) \sin M\phi
\end{align}\end{subequations}
As previously mentioned the frame index on the angles is not indicated as the formulas are identical in both the helicity and the GJ frames. However it is important to remember that the moments depend on the frame. The relation between the moments extracted in the helicity and the GJ frames is
\begin{align} \label{eq:rot_mom}
\left. H^\alpha(LM)\right|_\text{GJ} & = \sum_{M'} \left. H^\alpha(LM')\right|_\text{H} d^L_{MM'}(\omega),  & \text{with }\quad \alpha & = 0,1,2.
\end{align}
This equation is not limited to the helicity and the GJ frames but provide the relation between moments in any two frames related by a rotation around the $y$ axis, with $\omega$ being the angle between the frames. 
In principle, moments can be extracted for arbitrarily large angular momenta $L$. In practical cases, the angular distribution of a two mesons system up to an invariant mass $m_\text{max}$ can be accurately described with a finite number of moments $L \leqslant L_\text{max}(m_\text{max})$. Eq.~\eqref{eq:rot_mom} tells us that $L_\text{max}$ is the same in both frames. For a fixed $L_\text{max}$, the number of relevant moments in a given frame might be limited to specific $M$ values for dynamical reason. For instance, it may happen that, the helicity frame at high energy, $s$-channel helicity conservation restricts the relevant moments to only $M=1$. But even in this case Eq.~\eqref{eq:rot_mom} indicates that in any other frame, moments with all $M$ components will be contribute to the angular distribution. 

%-------------------------------------------------------------------------------------------------
\section{partial waves}
The quantum numbers of the $P_1P_2$ system are obtained from its partial wave decomposition. 
The partial wave amplitudes  $T^\ell_{\lambda m; \lambda_1\lambda_2}$ are defined by 
\begin{align}
A_{\lambda; \lambda_1\lambda_2} (\Omega)  =  \sum_{\ell m} T^\ell_{\lambda m; \lambda_1\lambda_2} Y^{m}_\ell(\Omega).
\end{align}
Furthermore it is convenient to work in the so-called reflectivity basis which uses the following linear combination of the two, $\lambda_\gamma = \pm 1$ photon helicities  \begin{align}  \label{def:Teps2}
^{(\epsilon)}T^\ell_{m; \lambda_1\lambda_2} & 
\equiv   \frac{1}{2} \left[ T^\ell_{+1 m; \lambda_1\lambda_2}  - \epsilon \: (-1)^m\ T^\ell_{-1 -m; \lambda_1\lambda_2}\right],
\end{align}
with $m = -\ell, \cdots,  \ell$. 
As shown, in Ref.~\cite{Mathieu:2019fts}, in the high-energy limit the amplitudes with  $\epsilon = +1(-1)$
are dominated by $t$-channel exchanges with naturality, $\eta=+1(-1)$, respectively. The reflectivity $\epsilon$ is the eigenvalue of the reflectivity operator, the symmetry through the reaction plane, and the naturality is defined by $\eta = P(-1)^J$ for exchanges of spin $J$ and parity $P$.
Parity invariance implies
\begin{align} \label{eq:parity2}
^{(\epsilon)}T^\ell_{m; -\lambda_1-\lambda_2} & = \epsilon (-1)^{\lambda_1-\lambda_2}\  ^{(\epsilon)}T^\ell_{m; \lambda_1\lambda_2}.
\end{align}
The target being a nucleon, there are only two possibilities for its helicity $\lambda_1 = \pm\frac{1}{2}$. 
There are thus $2\times (2\ell+1) \times (2s_R+1)$ independent partial wave amplitudes of spin $\ell$, $s_R$ being the spin of the recoiling particle. A standard choice is the set $\,  ^{(\epsilon)}T^\ell_{m; \frac{1}{2} \lambda_2}$. We designate these amplitudes by $[\ell]^{(\epsilon)}_{m;k}$ with $[\ell] = S, P, D, \ldots$, $k = 0,1,\ldots, 2 s_R+1$, $m = -\ell, \ldots, \ell$ and $\epsilon = \pm$. It is worth noting that in photoproduction, the reflectivity basis involves positive and negative values of $m$ while in the case of a spinless beam only the $m \geqslant 0$ spin projections enter~\cite{Chung:1974fq}. The reflectivity basis defined by Eq.~\eqref{def:Teps2} becomes more convenient at high energies, when the production mechanism is dominated by natural parity Reggeons. In this case, the set of relevant partial waves can be reduced by noting that $[\ell]_{m;k}^{(-)}  \ll [\ell]_{m;k}^{(+)}$.

Parity conservation implies that the two reflectivity components add incoherently in observables:
%Thanks to parity conservation the summation over reflectivity component is incoherent in observables:
\begin{align}
\sum_{\lambda_1,\lambda_2} \  ^{(\epsilon)}T^\ell_{m; \lambda_1\lambda_2} \  ^{(\epsilon')}T^{\ell'}_{m'; \lambda_1\lambda_2} & = 2 \delta_{\epsilon,\epsilon'} \sum_k [\ell]^{(\epsilon)}_{m;k} [\ell']^{(\epsilon)*}_{m';k}.
\end{align}
Therefore all observables can be decomposed into their reflectivity components:\begin{align}
I^\alpha(\Omega) &= \ ^{(+)}I^\alpha(\Omega)  + \ ^{(-)}I^\alpha(\Omega)  & 
H^\alpha(LM) & = \ ^{(+)}H^\alpha(LM) + \ ^{(-)}H^\alpha(LM) &
\alpha & = 0,1,2
\end{align}
The expressions of the moments up to $L=4$ in terms of $S$, $P$ and $D$ waves are provided in the Appendix. 

There are several equivalent ways of expressing the intensities in term of the partial waves in the reflectivity basis. One possibility is to introduce the quantities $Z_\ell^m(\Omega,\Phi) = Y_\ell^m(\Omega) e^{-i \Phi}$ such that
\begin{align}
\re Z_\ell^m(\Omega,\Phi) & = \sqrt{\frac{2\ell+1}{4\pi}} d^\ell_{m0}(\theta) \cos(m\phi-\Phi), &
\im Z_\ell^m(\Omega,\Phi) & = \sqrt{\frac{2\ell+1}{4\pi}} d^\ell_{m0}(\theta) \sin(m\phi-\Phi).
\end{align}
With this definition, the full intensity becomes
\begin{align} \nonumber
I(\Omega,\Phi)  = 2\kappa \sum_{k} \bigg\{ & (1+ P_\gamma) \left |\sum_{\ell,m} [\ell]_{m;k}^{(+)} \re Z_\ell^m (\Omega,\Phi) \right |^2 + 
(1- P_\gamma) \left |\sum_{\ell,m} [\ell]_{m;k}^{(+)} \im Z_\ell^m (\Omega,\Phi) \right |^2  \\
 + & (1- P_\gamma) \left |\sum_{\ell,m} [\ell]_{m;k}^{(-)} \re Z_\ell^m (\Omega,\Phi) \right |^2 + (1+P_\gamma)\left |\sum_{\ell,m} [\ell]_{m;k}^{(-)} \im Z_\ell^m (\Omega,\Phi) \right |^2 \bigg\}
\label{eq:all_in_one}
\end{align}
This expression can be used to extract partial waves $[\ell]^{(\epsilon)}_{m;k}$ from the measured intensity $I(\Omega,\Phi)$. %Without further input, the partial waves extracted this way
In practice the phase space is absorbed in a redefinition of the partial wave $\sqrt{\kappa} [\ell]_{m;k}^{(\epsilon)} \to [\ell]_{m;k}^{(\epsilon)}$ and, if nor the target or the recoiling particle is polarized, the summation over $k$ is ignored.

%-------------------------------------------------------------------------------------------------
\section{Beam asymmetries}
A common observable extracted with a polarized beam is the beam asymmetry, 
defined as the difference in the intensity between polarization parallel $\Phi = 0$ and perpendicular $\Phi = \frac{\pi}{2}$ to the reaction plane, normalized to their sum. When two mesons are produced, the decay angles of one of the meson $\Omega = (\theta,\phi)$ have to be specified. In Ref.~\cite{Mathieu:2019fts}, we provided a general definition of  beam asymmetries. We also detailed the properties of two special cases: the integrated beam asymmetry $\Sigma_{4\pi}$ and the beam asymmetry along the $y$ axis $\Sigma_y$. Both $\Sigma_{4\pi}$ and $\Sigma_y$ are independent of the $z$ axis and are equal in both the helicity and the GJ frames. $\Sigma_{4\pi}$ is obtained by integrating over the decay angles:
\begin{align} \label{eq:defBA}
\Sigma_{4\pi} &= \frac{1}{P_\gamma} \frac{\int_{4\pi} \left[I(\Omega,0) -  I(\Omega,\frac{\pi}{2}) \right] \diff\Omega}{\int_{4\pi} \left[I(\Omega,0) +  I(\Omega,\frac{\pi}{2}) \right] \diff\Omega}
= \frac{-1}{P_\gamma} \frac{\int_{4\pi} I^1(\Omega)\diff \Omega}{\int_{4\pi}I^0(\Omega) \diff\Omega}  =\frac{H^1(00)}{H^0(00)}. 
\end{align}
$\Sigma_{4\pi}$ has the advantage of being easily extracted experimentally but its interpretation in term of partial waves is not obvious in general, as can be seen from
\begin{align}
H^0(00) & = 2\kappa\sum_{k,\ell,m} \left|[\ell]_{m;k}^{(+)}\right|^2 + \left|[\ell]_{m;k}^{(-)}\right|^2 , &
H^1(00) & = 2\kappa\sum_{k,\ell,m} (-1)^m \re\left([\ell]_{m;k}^{(+)} [\ell]_{-m;k}^{(+)*} - 
[\ell]_{m;k}^{(-)} [\ell]_{-m;k}^{(-)*}\right).
\end{align} 
Although positive an negative reflectivity components contribute with opposite signs, the terms $ \re \left( [\ell]_{m;k}^{(\epsilon)} [\ell]_{-m;k}^{(\epsilon)*} \right)$ are not positive definite. There is thus no relation between the sign of $\Sigma_{4\pi}$ are the reflectivity of the waves.

The beam asymmetry $\Sigma_y$ in which the two meson momenta were perpendicular to the reaction plane is given by
\begin{align}\label{eq:defBAy}
\Sigma_y & = \frac{1}{P_\gamma} \frac{I(\Omega_y,0) - I(\Omega_y,\frac{\pi}{2})}{I(\Omega_y,0) + I(\Omega_y,\frac{\pi}{2})} =  -\frac{I^1(\Omega_y)}{I^0(\Omega_y)}~.
\end{align}
Binning around the $y$ axis, its angles are $\Omega_y = (\frac{\pi}{2}, \frac{\pi}{2})$, drastically reduce the number of events. Alternatively, $\Sigma_y$ is advantageously reconstructed from the moments. 
The expression of the intensities $I^\alpha(\Omega_y)$ with $\alpha = 0,1$ in terms of moments, truncated to $L=4$, is
\begin{align}% \nonumber
4\pi(-1)^\alpha I^\alpha(\Omega_y) & = H^\alpha(00) - \frac{5}{2} H^\alpha(20) - 5 \sqrt{\frac{3}{2}} H^\alpha(22) + \frac{27}{8} H^\alpha(40) + \frac{9}{2}\sqrt{\frac{5}{2}} H^\alpha(42) + \frac{9}{4}\sqrt{\frac{35}{2}}  H^\alpha(44). \label{eq:momy}
\end{align}
$\Sigma_y$ is designed to represent the action of the reflection through the reaction plane. By construction, the production of the wave $[\ell]^{(\epsilon)}_{m;k}$ has the reflectivity eigenvalue $\epsilon$ and the decay amplitude has the eigenvalue $(-1)^\ell$ when the meson momenta are aligned with the $y$ axis. In order words, a single partial wave $[\ell]^{(\epsilon)}_{m;k}$ leads to $\Sigma_y = \epsilon (-1)^\ell$.  When only one production mechanism dominates, as expected at high energies, $\Sigma_y$ provide information about the parity of the dominant waves in a given invariant mass region. This observables might thus be useful to detect exotic waves. In the case of two identical mesons, only even waves contributes to the system by Bose symmetry. In this case $\Sigma_y$ provides an information about the relative weight of each reflectivity component, as it is in single pseudoscalar photoproduction. For completeness, the expression of the intensities involved in $\Sigma_y$ in the presence of multiple waves is given by
\bsub \label{eq:I0I1}
\begin{align}
I^0(\Omega_y) & =2 \kappa \sum_{k,\ell, \ell'} \phantom{(-1)^\ell} \sum_{m, m'} \left([\ell]^{(+)}_{m;k} [\ell']^{(+)*}_{m';k} + [\ell]^{(-)}_{m;k} [\ell']^{(-)*}_{m';k} \right)Y_\ell^m(\Omega_y) Y_{\ell'}^{m'*}(\Omega_y)
\\
-I^1(\Omega_y) & = 2 \kappa \sum_{k, \ell, \ell'} (-1)^\ell \sum_{m,m'} \left([\ell]^{(+)}_{m;k} [\ell']^{(+)*}_{m';k} - [\ell]^{(-)}_{m;k} [\ell']^{(-)*}_{m';k} \right) Y_\ell^m(\Omega_y)Y_{\ell'}^{m'*}(\Omega_y),
\end{align}
\esub
where the sums are restricted to $\ell, \ell', m, m'$ having the same parity.
For completeness, we mention that 
\begin{align}
Y_{\ell}^{m}(\Omega_y) &= i^\ell \sqrt{\frac{2\ell+1}{4\pi}} \sqrt{ \frac{(\ell-m)!}{(\ell+m)!}} \frac{(\ell+m-1)!!}{(\ell-m)!!},&
\text{for even} \quad \ell+m.
\end{align}

%\section{Conclusion}

\begin{acknowledgments}
I acknowledges support from the Community of Madrid through 
the Programa de atracción de talento investigador 2018 (Modalidad 1).
\end{acknowledgments}

%\newpage
\appendix
%******************************************************************************************
\section{\boldmath Moments with $S$, $P$ and $D$ waves} \label{sec:app}
I restrict the wave set to only $S$, $P$ and $D$ waves and ignore the $s_R$ degrees of freedom relative to the recoiling particles. I also suppress the phase space factor $\kappa$ in the formulas below.
There are $1+3+5$ waves for each reflectivity components. Since moments involve an incoherent sum over the two reflectivity components, one just needs to specify the relation between moments and one reflectivity component of the partial waves. In the following, only the relation involving the positive reflectivity components are provided for simplicity and one should keep in mind that the moments are obtained by summing both reflectivity components using
\bsub\begin{align}
H^{0}(LM) &=\ ^{(+)}H^{0}(LM) +\  ^{(-)}H^{0}(LM) \\
H^{1}(LM) &=\ ^{(+)}H^{1}(LM) +\  ^{(-)}H^{1}(LM) \\
\im H^{2}(LM) &=\ ^{(+)}H^{2}(LM) + \ ^{(-)}H^{2}(LM)
\end{align} \esub
The moments $\  ^{(-)}H^{0}(LM)$ are obtained from $\  ^{(+)}H^{0}(LM)$ by replacing the $\epsilon = +$ waves by the $\epsilon = -$ waves. The moments $\  ^{(-)}H^{1,2}(LM)$ are obtained from $\  ^{(+)}H^{1,2}(LM)$ by replacing the $\epsilon = +$ waves by the $\epsilon = -$ waves {\bf and by changing the overall sign}. A C/C++ implementation of the moments can be found on the JPAC website~\cite{JPACweb, Mathieu:2016mcy}. 
Their expressions are given below, classified according to the angular momentum $L$. 

$L = 0$:
\bsub\begin{align}
^{(+)}H^{0}(00) &= 2 \left[  |\Sz|^2  +|P_{-1}^{(+)}|^2 + |P_0^{(+)}|^2 +  |P_1^{(+)}|^2 + |D_{-2}^{(+)}|^2  + |D_{-1}^{(+)}|^2 + |D_0^{(+)}|^2   + |D_1^{(+)}|^2  + |D_2^{(+)}|^2   \right]~, \\
^{(+)}H^{1}(00) &= 2 \left[ |S_0^{(+)}|^2  + |P_0^{(+)}|^2 + |D_0^{(+)}|^2 - 2 \re (\Pp \Pms)  - 2 \re \left(\Dp \Dms \right)  - 2 \re \left(\Dpp \Dmms \right)  \right]~
\end{align}\esub

$L = 1$:
\bsub\begin{align}
^{(+)}H^{0}(10) &= \frac{4}{\sqrt{3}} \re \left(\Sz \Pzs \right) + \frac{8}{\sqrt{15}} \re \left(\Pz \Dzs \right) + 
\frac{4}{\sqrt{5}} \re \left(\Pp \Dps \right) + \frac{4}{\sqrt{5}} \re \left(\Pm \Dms \right)\\
^{(+)}H^{1}(10) &=  \frac{4}{\sqrt{3}} \re \left(\Sz \Pzs \right) + \frac{8}{\sqrt{15}} \re \left(\Pz \Dzs \right) - \frac{4}{\sqrt{5}} \re\left(\Pp \Dms + \Pm \Dps \right)\\
\nonumber
^{(+)}H^{0}(11) &= \frac{2}{\sqrt{3}} \re\left[ \left(\Sz  - \frac{1}{\sqrt{5}} \Dz\right) \left(\Pps -\Pms \right) \right]  +  \frac{2}{\sqrt{5}} \re(\Pz \Dps - \Pz \Dms) \\
& +  \frac{2\sqrt{2}}{\sqrt{5}} \re(\Pp \Dpps - \Pm \Dmms) \\
\nonumber
^{(+)}H^{1}(11) &=  \frac{2}{\sqrt{3}} \re\left[ \left(\Sz  - \frac{1}{\sqrt{5}} \Dz\right) \left(\Pps -\Pms \right) \right]  +  \frac{2}{\sqrt{5}} \re(\Pz \Dps - \Pz \Dms)  \\
& +  \frac{2\sqrt{2}}{\sqrt{5}} \re(\Pp \Dmms - \Pm \Dpps)\\
\nonumber
^{(+)}H^{2}(11) &= -\frac{2}{\sqrt{3}} \re\left[ \left(\Sz  - \frac{1}{\sqrt{5}} \Dz\right) \left(\Pps +\Pms \right) \right] -  \frac{2}{\sqrt{5}} \re(\Pz \Dps + \Pz \Dms) \\
& +  \frac{2\sqrt{2}}{\sqrt{5}} \re(\Pp \Dmms + \Pm \Dpps)
\end{align}\esub

$L = 2$:
\bsub\begin{align} \nonumber
^{(+)}H^{0}(20) &= \frac{4}{5} |\Pz|^2 - \frac{2}{5} \left(|\Pp|^2 + |\Pm|^2 \right) + \frac{4}{7} |\Dz|^2 + \frac{2}{7} \left(|\Dp|^2 + |\Dm|^2 \right)  \\ 
& - \frac{4}{7} \left(|\Dpp|^2 + |\Dmm|^2 \right) + \frac{4}{\sqrt{5}} \re\left(\Sz\Dzs \right)\\
\nonumber
^{(+)}H^{1}(20) &= \frac{4}{5} |\Pz|^2 + \frac{4}{5} \re \left( \Pp \Pms \right) + \frac{4}{7} |\Dz|^2 - \frac{4}{7} \re \left( \Dp \Dms \right)  - \frac{8}{7} \re \left( \Dpp \Dmms \right) \\ 
& +  \frac{4}{\sqrt{5}} \re\left(\Sz\Dzs \right)\\
\nonumber
^{(+)}H^{0}(21) &= 2 \re \left[\left(\frac{\Sz}{\sqrt{5}} + \frac{ \Dz}{7} \right)\left(\Dps-\Dms \right) \right]  + 2\frac{\sqrt{3}}{5} \re\left[\Pz \left( \Pps-\Pms\right) \right]  \\
& + 2\frac{\sqrt{6}}{7} \re\left[\Dp \Dpps - \Dm\Dmms \right] \\
\nonumber
^{(+)}H^{1}(21) &=  2 \re \left[\left(\frac{\Sz}{\sqrt{5}} + \frac{ \Dz}{7} \right)\left(\Dps-\Dms \right) \right]+ 2\frac{\sqrt{3}}{5} \re\left[\Pz \left( \Pps-\Pms\right) \right]\\
& +  2\frac{\sqrt{6}}{7} \re\left[\Dp \Dmms + \Dm\Dpps\right] \\
^{(+)}H^{2}(21) &= -2 \re \left[\left(\frac{\Sz}{\sqrt{5}} + \frac{ \Dz}{7} \right)\left(\Dps+\Dms \right) \right] - 2\frac{\sqrt{3}}{5} \re\left[\Pz \left( \Pps+\Pms\right) \right]\\
& +  2\frac{\sqrt{6}}{7} \re\left[\Dp \Dmms + \Dm\Dpps\right]  
\\
%\end{align}\esub
%\bsub\begin{align} 
^{(+)}H^{0}(22) &= 2 \re \left[\left(\frac{\Sz}{\sqrt{5}} - \frac{2}{7} \Dz\right)\left(\Dpps+\Dmms \right) \right] - 2 \frac{\sqrt{6}}{5} \re \left(\Pp \Pms \right) - 2 \frac{\sqrt{6}}{7} \re \left(\Dp \Dms \right)\\
^{(+)}H^{1}(22) &= 2 \re \left[\left(\frac{\Sz}{\sqrt{5}} - \frac{2}{7} \Dz\right)\left(\Dpps+\Dmms \right) \right]+ \frac{\sqrt{6}}{5} \left(|\Pp|^2  + |\Pm|^2\right)  + \frac{\sqrt{6}}{7} \left(|\Dp|^2  + |\Dm|^2\right)  \\
^{(+)}H^{2}(22) &= -2 \re \left[\left(\frac{\Sz}{\sqrt{5}} - \frac{2}{7} \Dz\right)\left(\Dpps-\Dmms \right) \right]- \frac{\sqrt{6}}{5} \left(|\Pp|^2  - |\Pm|^2\right)  - \frac{\sqrt{6}}{7} \left(|\Dp|^2  - |\Dm|^2\right) 
\end{align}\esub

$L = 3$:
\bsub\begin{align}
^{(+)}H^{0}(30) &= \frac{12}{7\sqrt{5}} \re \left(\sqrt{3} \Pz \Dzs - \Pp\Dps - \Pm \Dms \right)\\
^{(+)}H^{1}(30) &= \frac{12}{7\sqrt{5}} \re \left(\sqrt{3} \Pz \Dzs + \Pp\Dps + \Pm \Dms \right)
\\
\nonumber
^{(+)}H^{0}(31) &= \frac{4}{7} \sqrt{\frac{6}{5}} \re \left[\Pz \left(\Dps- \Dms \right) \right] + \frac{6}{7} \sqrt{\frac{2}{5}} \re \left[\Dz \left(\Pps- \Pms \right) \right] \\
& - \frac{2}{7} \sqrt{\frac{3}{5}} \re \left(\Pp \Dpps- \Pm\Dmms \right) \\
\nonumber
^{(+)}H^{1}(31) &=  \frac{4}{7} \sqrt{\frac{6}{5}} \re \left[\Pz \left(\Dps- \Dms \right) \right] + \frac{6}{7} \sqrt{\frac{2}{5}} \re \left[\Dz \left(\Pps- \Pms \right) \right] \\
& - \frac{2}{7} \sqrt{\frac{3}{5}} \re \left(\Pp \Dmms- \Pm\Dpps \right) 
%\\
\end{align}\esub
\bsub\begin{align} 
\nonumber
^{(+)}H^{2}(31) &=  -\frac{4}{7} \sqrt{\frac{6}{5}} \re \left[\Pz \left(\Dps+ \Dms \right) \right] - \frac{6}{7} \sqrt{\frac{2}{5}} \re \left[\Dz \left(\Pps+ \Pms \right) \right] \\
& - \frac{2}{7} \sqrt{\frac{3}{5}} \re \left(\Pp \Dmms+ \Pm\Dpps \right)
\\
^{(+)}H^{0}(32) &= \frac{\sqrt{12}}{7} \re \left[\Pz \left( \Dpps + \Dmms \right) - \sqrt{2}\Pp\Dms -\sqrt{2} \Pm \Dps \right)\\
^{(+)}H^{1}(32) &= \frac{\sqrt{12}}{7} \re \left[\Pz \left( \Dpps + \Dmms \right) + \sqrt{2}\Pp\Dps +\sqrt{2} \Pm \Dms \right)\\
^{(+)}H^{2}(32) &= - \frac{\sqrt{12}}{7} \re \left[\Pz \left( \Dpps - \Dmms \right) + \sqrt{2}\Pp\Dps -\sqrt{2} \Pm \Dms \right)\\
^{(+)}H^{0}(33) &= \frac{6}{7} \re \left[\Pz  \Dmms -\Pm \Dpps \right)\\
^{(+)}H^{1}(33) &= \frac{6}{7} \re \left[\Pz  \Dpps -\Pm \Dmms \right)\\
^{(+)}H^{2}(33) &= \frac{6}{7} \re \left[\Pz  \Dpps +\Pm \Dmms \right)
\end{align}\esub

$L = 4$:
\bsub\begin{align}
^{(+)}H^{0}(40) &= \frac{2}{21} \left[ 6 |\Dz|^2 - 4 \left( |\Dp|^2 + |\Dm|^2 \right) + |\Dpp|^2 + |\Dmm|^2\right]\\
^{(+)}H^{1}(40) &=  \frac{4}{21} \left[ 3 |\Dz|^2 +  \re \left( 4 \Dp\Dms + \Dpp\Dmms \right) \right]\\
^{(+)}H^{0}(41) &= \frac{2 \sqrt{5}}{21} \re \left[ \sqrt{6} \Dz \left(\Dps-\Dms \right)  + \Dm\Dmms - \Dp \Dpps \right]\\
^{(+)}H^{1}(41) &= \frac{2 \sqrt{5}}{21} \re \left[ \sqrt{6} \Dz \left(\Dps-\Dms \right)  + \Dm\Dpps - \Dp \Dmms \right]\\
^{(+)}H^{2}(41) &= -\frac{2 \sqrt{5}}{21} \re \left[ \sqrt{6} \Dz \left(\Dps-\Dms \right)  + \Dm\Dpps + \Dp \Dmms \right]\\
^{(+)}H^{0}(42) &= \frac{\sqrt{10}}{21} \re \left[ \sqrt{6} \Dz \left(\Dpps+\Dmms \right)  - 4 \Dp\Dms  \right]\\
^{(+)}H^{1}(42) &= \frac{\sqrt{10}}{21} \re \left[ \sqrt{6} \Dz \left(\Dpps+\Dmms \right)  + 2 |\Dp|^2  + 2 |\Dms|^2  \right]\\
^{(+)}H^{2}(42) &= - \frac{\sqrt{10}}{21} \re \left[ \sqrt{6} \Dz \left(\Dpps-\Dmms \right)  + 2 |\Dp|^2  - 2 |\Dms|^2  \right]\\
^{(+)}H^{0}(43) &= \frac{2}{3}\sqrt{\frac{5}{7}} \re \left[ \Dp \Dmms- \Dm\Dpps \right]\\
^{(+)}H^{1}(43) &=  \frac{2}{3}\sqrt{\frac{5}{7}} \re \left[ \Dp \Dpps- \Dm\Dmms \right]\\
^{(+)}H^{2}(43) &= -\frac{2}{3}\sqrt{\frac{5}{7}} \re \left[ \Dp \Dpps+ \Dm\Dmms \right]\\
^{(+)}H^{0}(44) &= \frac{2}{3} \sqrt{\frac{10}{7}} \re \left[ \Dpp \Dmms \right]\\
^{(+)}H^{1}(44) &=\frac{1}{3} \sqrt{\frac{10}{7}}  \left[ |\Dpp|^2+ |\Dmms|^2 \right] \\
^{(+)}H^{2}(44) &= - \frac{1}{3} \sqrt{\frac{10}{7}} \left[ |\Dpp|^2- |\Dmms|^2 \right]
\end{align}\esub

\end{document}